\documentclass[useAMS,usenatbib]{mn2e}
\usepackage{graphicx}

\usepackage{color}

\title[Emission from star-planet collisions]
{Optical-infrared flares and radio afterglows 
by Jovian planets inspiraling into their host stars
} 
\author[Yamazaki, Hayasaki, and Loeb]
{Ryo Yamazaki$^{1,2}$\thanks{E-mail: ryo@phys.aoyama.ac.jp},
Kimitake~Hayasaki$^{2,3}$
and
Abraham Loeb$^{2}$
\\
$^{1}$Department of Physics and Mathematics, Aoyama Gakuin University, 
5-10-1, Fuchinobe, Sagamihara 252-5258, Japan \\
$^{2}$Harvard-Smithsonian Center for Astrophysics, 60 Garden Street, Cambridge, MA, 02138, USA
\\
$^{3}$Department of Astronomy and Space Science, Chungbuk National University, Cheongju
361-763, Korea
}

\begin{document}


\pagerange{\pageref{firstpage}--\pageref{lastpage}} \pubyear{2015}
\maketitle
\label{firstpage}


\begin{abstract}
When a planet inspirals into its host star,
it releases gravitational energy which is converted into
an expanding bubble of hot plasma.
We study the  radiation from  the bubble and show that it includes
prompt optical-infrared emission and a subsequent radio afterglow.
The prompt emission from M31 and Large Magellanic Cloud is 
detectable by optical-near infrared transient surveys with a large 
field of view. The subsequent 
radio afterglows are detectable for $10^{3-4}$~years.
The event rate depends on  uncertain parameters in the
formation and dynamics of giant planets.
Future observation of the rate will constrain related theoretical models.
 If the event rate is high ($\ga$~a few events per year),
 the circumstellar disk must typically be massive as suggested by recent numerical simulations.

\end{abstract}


%
\section{Introduction}
%

A substantial fraction of gaseous planetary-mass objects might 
be ingested by the central stars \citep[][and references therein]{Machida2010,Machida2011,Inutsuka2012,Vorobyov2010,Vorobyov2015}.
Realistic non-ideal magnetohydrodynamics simulations have shown that
protoplanetary disks are initially massive enough to produce multiple
Jupiter-mass planets via gravitational instability 
\citep{Inutsuka2010,Machida2011,Tsukamoto2015}. 
These massive planets may  survive in the subsequent era, during 
which planets gravitationally interact or collide with each other to 
produce hot Jupiters and highly eccentric planets
\citep[e.g.,][]{Ida2004,Chatterjee2008,Ford2008}.
A large fraction of the hot Jupiters which migrate to the vicinity 
of the central stars are either consumed \citep{Sandquist1998} 
or tidally disrupted \citep{Gu2003} by the host star. Stars without 
detected hot Jupiters might have already ingested them 
\citep{Rice2008,Inutsuka2009,Ogihara2013,Ogihara2014}.
Present-day hot Jupiters could secularly enlarge their eccentricity 
to reach their host stars by a process like the Kozai-Lidov mechanism 
\citep{Kozai1962,Lidov1962}. Several ways for detecting stellar 
ingestion of planets have been proposed 
\citep[e.g.,][]{Sandquist1998,Sandquist2002,Cody2005,Jackson2009,Teitler2014,Matsakos2015}.
If a star ingests planets on average $N_{\rm i}$ times during its life, 
the total event rate in the Galaxy is estimated to be
${\rm SFR}\times N_{\rm i}/\langle m\rangle\sim5N_{\rm i}$~yr$^{-1}$,
where ${\rm SFR}\approx1 M_{\sun}$~yr$^{-1}$ is the star formation rate of 
the Milky Way \citep{Robitaille2010} 
and $\langle m\rangle \approx0.2M_{\sun}$ is the average stellar mass
\citep{Kroupa2001,Chabrier2003}.

In this paper, we calculate the radiation expected at the moment of 
Jovian planet ingestion and the subsequent afterglow phase. 
When a planet is engulfed by a host star,
it releases gravitational energy which is converted into
an expanding bubble of hot plasma (\S~2).
The expanding bubble generates an optical-infrared flare (\S~3). 
Subsequently, the material interacts with the  circumstellar matter (CSM),
generating a shock that accelerates electrons to relativistic energies,
which in turn produce synchrotron radiation (\S~4).
There are several previous papers that predicted transient emissions from planet-star
interaction.
\citet{Metzger2012} focused on the case of quasi-circular orbit of the hot Jupiters
around a main sequence star, which produces super-Eddington accretion.
\citet{Bear2011} studied tidal disruption of planets by brown dwarfs.
These authors argued that the accretion disk drives outflows,
producing long-duration ($>{\rm day}$) transients.
On the other hand, we argue that a planet with a highly eccentric orbit
in-spirals into the star, depositing thermal energy near the stellar surface,
which produces the expanding plasma bubble.

\section{Interaction between a star and a Jovian planet}
\label{sec:ingestion}
%
Most extrasolar planets reside in eccentric orbits \citep{Marcy2005,Winn2015}. 
Some planets are subject to rapid orbital change or migration by the interaction 
with a gaseous disk \citep{Rice2008}, planet-planet scattering 
\citep{Rasio1996,Weidenschilling1996}, or the Kozai-Lidov mechanism 
\citep{Ford2000,Fabrycky2007,Li2015}. 
Following these dynamical processes, the pericentre distance 
$R_{\rm p}$ finally becomes shorter than the tidal disruption 
radius of a planet: $R_{\rm T}=(M_*/m_{\rm pl})^{1/3}r_{\rm pl}$, 
where $M_*$ is a mass of the star hosting the planet with a mass 
$m_{\rm pl}$ and a radius $r_{\rm pl}$. 
Then, the planet is tidally disrupted 
(Rees~1988; Phinney~1989; Evans \& Kochanek~1989 for 
the context of a supermassive black hole system, Faber et al.~2005; 
Guillochon et al. 2011 for a star-planet system). 
Note that $R_T$ is as large as the radius of the star $R_*$ for Sun-like G-type 
stars and Jovian mass planets, whereas $R_T>R_*$ for K- and M-stars \cite[e.g.,][]{Rappaport2013}.
If $R_{\rm p}$ is smaller than $R_{\rm T}$ for the Sun-like G-type star 
we consider in this paper, the approaching planet can impact directly the stellar surface. 
In other words, if a penetration factor $\beta=R_{\rm T}/R_{\rm p}=1$, the planet 
is tidally disrupted by the host star. Otherwise, for $\beta>1$ the planet would collide 
with the host star. Next, we will discuss these two cases separately
in the following subsections.

%
\subsection{Tidal disruption of a planet by a star:\,$\beta=1$}
%

After the tidal disruption of a planet on a parabolic orbit, the debris mass is distributed around zero specific energy (e.g., see Figure 3 of \citealt{Evans1989}), characteristic of a parabolic orbit. If the planet originally approaches the star on a bound (eccentric or circular) orbit, however, the mass distribution of the disrupted planet would shift to negative specific energy. This gives the condition to cause tidal disruption of planets on eccentric orbits,  $0\le{e}<e_{\rm crit}$, where 
$e_{\rm crit}$ is related to a penetration factor $\beta$ \citep{Hayasaki2013}.
\begin{eqnarray}
e_{\rm crit}=1-\left(\frac{2}{\beta}\right)\left(\frac{m_{\rm pl}}{M_{*}}\right)^{1/3}~~.
\label{eq:crit}
\end{eqnarray}
 On the other hand, $e_{\rm crit}<{e}\,{\le}1$ must be satisfied for the parabolic 
 tidal disruption events (TDEs). In order for the planet to be tidally disrupted, the pericentre distance must be smaller than the tidal disruption radius. This gives a constraint on the orbital eccentricity:
\begin{eqnarray}
e\ge 1-\frac{R_{\rm T}}{a}
\label{eq:econd}
\end{eqnarray}
Combining Eqs.~(\ref{eq:crit}) with (\ref{eq:econd}), we can obtain the alternative condition that $a_{\rm crit}<a<\infty$ for parabolic TDEs and $0<a<a_{\rm crit}$ for the eccentric TDEs, where 
\begin{eqnarray}
a_{\rm crit}=\frac{\beta}{2}\left(\frac{M_{*}}{m_{\rm pl}}\right)^{1/3}R_{\rm T}.
\end{eqnarray}
For the Sun-like G-type stars with Jovian planets, the critical semi-major axis is estimated to be $a_{\rm crit}\approx5\beta{R}_\odot$, so that parabolic TDEs only occur if $a\ga{a}_{\rm crit}$ for $\beta=1$.
The eccentric TDEs are likely to happen in the case of the system composing of a star and a hot-jupiter on a quasi-circular orbit, because of $0<a\la{a_{\rm crit}}$\footnote{The tidal disruption of the planet given by Section 5 of \citet{Metzger2012} corresponds to the eccentric TDEs we have proposed here.}.

Assuming the full conservation of angular momentum during debris circularization, 
the circularization radius is given by $r_{\rm c}=(R_{\rm T}/\beta)(1+e)$ for 
Keplerian rotation. If $\beta=1$, an accretion disc forms around the star 
at $R_{\rm *}\la{r}_{\rm c}\la 2R_{*}$ for $0\la{e}\la1$. 
Because of the super-Eddington accretion nature of mass fallback rate, 
the super-Eddignton outflow is likely to be caused \citep{Strubbe2009}. 
This outflow would produce the long-duration ($\gg {\rm day}$) optical 
transients occurred in the star-planet system \citep{Metzger2012}. 
On the other hand, if $\beta > 1$, the planet impacts the stellar surface and is 
subject to orbital friction within the stellar envelope. In this regime, 
it is not clear whether the tidal disruption occurs.

%
\subsection{Direct collision between a star and a planet:\,$\beta>1$}
\label{sec:collision}
\subsubsection{Simple order-of-magnitude estimate}
\label{subsec:collision}
%

Let us consider the situation that the planet on highly eccentric orbit hits 
the star. The ram pressure the planet gets from the star 
is given by $\rho_*v_{\rm p}^2$, 
where $\rho_{\rm *}=(3/4\pi)M_{*}/R_*^3$
and $v_{\rm p}=\sqrt{GM_*/R_{*}}$ are the average stellar density and 
planetary velocity, respectively. 
Then,  the drag force on  the planet is given by, 
\begin{equation}
F_{\rm drag}=\eta\sigma_{\rm p}\rho_{*}v_{\rm p}^2, 
\label{eq:dragforce}
\end{equation}
where $\eta=\mathcal{O}(1)$ and $\sigma_{\rm p}$ are the geometrical parameter 
and the cross section, respectively. For simplicity, we adopt $\sigma_{\rm p}=\pi{r_{\rm pl}^2}$.
Our estimated drag force is in agreement with recent simulation results
\citep{Staff2016}.
If the planet travels a distance $d$ inside the star, the total energy lost by the drag force 
$\Delta{E}_{\rm p}$ can be estimated as,
\begin{equation}
\Delta{E}_{\rm p}=F_{\rm drag}\cdot{d}
=\frac{3}{4}\eta\frac{GM_*^2}{R_*}
\left(\frac{r_{\rm pl}}{R_{*}}\right)^2
\left(\frac{d}{R_*}\right)~~.
\end{equation}

The kinetic energy of the planet at the periastron is given by, 
\begin{eqnarray}
E_{\rm p}
&=&\frac{1}{2}m_{\rm pl}v_{\rm peri}^2\sim\frac{\beta}{2}\frac{GM_*m_{\rm pl}}{R_*}
\nonumber \\
&=&3.6\times10^{45}
\left(\frac{\beta}{2}\right)
\left(\frac{\xi_*}{1}\right)
\left(\frac{m_{\rm pl}}{m_{\rm J}}\right)
\,{\rm erg},
\end{eqnarray}
where $v_{\rm peri}=\sqrt{GM_*/R_{\rm p}}\sim\sqrt{\beta GM_*/R_*}$ 
is the planetary velocity at the pericenter, 
$\xi_*=(M_*/M_{\sun})/(R_*/R_{\sun})$ and $m_{\rm J}=1.898\times10^{30}$g 
is Jupiter\rq{}s mass. Note that $\xi_*$ is of order  unity for main sequence stars with $M_*\la 2M_{\sun}$
\citep[e.g.,][]{Torres2010,Eker2015}.

The ratio of $\Delta{E_{\rm p}}$ and $E_{\rm p}$ is given by
\begin{eqnarray}
\frac{\Delta{E}_{\rm p}}{E_{\rm p}}=\frac{3}{2}\left(\frac{\eta}{\beta}\right)
\left(\frac{M_*}{m_{\rm pl}}\right)\left(\frac{r_{\rm pl}}{R_*}\right)^2
\left(\frac{d}{R_*}\right)~~.
\label{eq:eratio}
\end{eqnarray}
For a  Sun-like G-type star with the Jovian planet
and $\eta\sim1$ and $\beta\sim2$, 
$\Delta{E}_{\rm p}/E_{\rm p}\sim10$ for $d=R_*$. 
This suggests that the planet spirals into the star and stops 
when $\Delta{E}_{\rm p}\approx E_{\rm p}$
and $d\approx0.1R_*\sim r_{\rm J}$,
where $r_{\rm J}=7.0\times10^9$~cm is the radius of Jupiter.
Following the constraint on $\beta$ for $d<r_{\rm peri}$ 
as $2<\beta<\sqrt{15}$ from Eq.~(\ref{eq:eratio}), 
we find that for moderate values of $\beta$ the inspiraling planet could stop in the stellar 
external layer before causing the tidal disruption.
The friction between the planet and the stellar gas deposits the thermal energy.
If the supersonic planet motion makes a bow shock around the planet,
dissipated energy also turns into the thermal energy.
In these ways, independently of whether the tidal disruption occurs, 
the thermal energy,
\begin{equation}
E_{\rm th,i}\sim\frac{GM_*m_{\rm pl}}{2R_*}=1.8\times10^{45}\xi_*
\left(\frac{m_{\rm pl}}{m_{\rm J}}\right)~{\rm erg}~~,
\label{eq:initialEth}
\end{equation}
is injected at the stellar external layer.
The gas is optically thick, so that the radiative cooling is inefficient,
making the hot, adiabatically expanding plasma bubble.


\subsubsection{An example of star-planet collision: the case of
Sun-like star and a Jupiter-like planet. }
\label{subsec:example}
%
Below we consider the collision between a Sun-like star and a Jupiter-like
 giant planet
as a typical example. 
We start by deriving  simple power-law scaling relations for the solar interior. 
For a geometrically thin adiabatic gas with an adiabatic index $\gamma_g$,
hydrostatic equilibrium  gives the temperature as a function of the vertical depth $z$ 
(distance from the solar surface) as
$T_*(z)=\mu \mu_p(\gamma_g-1)gz/k_B\gamma_g$, where $g$, $\mu_{\rm p}$ and $\mu$
are the gravitational acceleration, the proton mass 
and the mean molecular weight, respectively.
In terms of the polytropic index $n=(\gamma_g-1)^{-1}$, we find that
the mass density and pressure scale as $\rho_*\propto z^n$ and
$p_*\propto z^{n+1}$.
Comparing these scalings with standard solar models \citep{Guenther1992}, 
we find that the simple power-law expressions with $n=2$ ($\gamma_g=3/2$),
\begin{eqnarray}
T_*(z)&=&1\times10^6~{\rm K}\left(\frac{z}{10^{10}{\rm cm}}\right)~~,
\label{eq:sun_interior1}\\
\rho_*(z)&=&0.05~{\rm g~cm}^{-3}\left(\frac{z}{10^{10}{\rm cm}}\right)^2~~,
\label{eq:sun_interior2}\\
p_*(z)&=&5\times10^{12}~{\rm dyn~cm}^{-2}\left(\frac{z}{10^{10}{\rm cm}}\right)^3~~,
\label{eq:sun_interior3}
\end{eqnarray}
approximate the solar interior well for $z\la 2\times10^{10}$cm.
Note that main sequence stars with a mass larger than $1 M_{\sun}$
are more concentrated than $n = 3$ polytropes
\citep[see, for example, Figure~A1 of][]{Freitag2005}. 

The deceleration of the planet inside a Sun-like star is described 
by\footnote{
For a similar study of a Jupiter-comet collision, see
\citet{Chevalier1994}.
}
\begin{equation}
\frac{{\rm d}\ln v_{\rm p}}{{\rm d}z} =
-\frac{\eta\rho_*(z)}{\rho_{pl}r_{pl}\cos\theta}~~,
\label{eq:deceleration}
\end{equation}
where 
$\rho_{pl}=3m_{pl}/4\pi r_{pl}^3$ is the average mass density of the planet,
$\eta$ is a numerical factor of order of unity [see Eq.~(\ref{eq:dragforce})], and
$\theta$ is an angle between the surface normal and the velocity of the planet 
just before the collision, that is, $z=d\cos\theta$.
With the initial velocity $v_{\rm p0}$ at $z=0$, 
equations~(\ref{eq:sun_interior2}) and (\ref{eq:deceleration}) give
\begin{equation}
v_{\rm p}(z)=v_{{\rm p}0}\exp \left[-\left(\frac{z}{z_s}\right)^3\right]~~,
\label{eq:planetvelocity}
\end{equation}
where the stopping depth $z_s$ is given by
\begin{equation}
z_s=3.8\times10^{10}~{\rm cm}~
\left(\frac{\cos\theta}{\eta}\right)^{1/3}
\left(\frac{m_{\rm pl}}{m_{\rm J}}\right)^{1/3}
\left(\frac{r_{\rm pl}}{r_{\rm J}}\right)^{-2/3}~~.
\label{eq:stopping}
\end{equation}
Hence the planet keeps its velocity until $z\la z_s$, and almost suddenly 
stops at  a depth $z_s$.
For a Sun-like density structure, the density more rapidly increases
 for $z\ga2\times10^{10}$~cm, so that
the rapid deceleration starts before $z_s$.
Note that in deriving  Eq.~(\ref{eq:stopping}), several effects were neglected for 
the purpose of a simple analytical calculation.
The sound speed of the stellar gas is
\begin{eqnarray}
c_{s*}(z)&=&\left(\frac{\gamma_g p_*(z)}{\rho_*(z)}\right)^{1/2}
\nonumber\\
&\approx& 1\times10^{7}~{\rm cm~s}^{-1}~
\left(\frac{z}{10^{10}{\rm cm}}\right)^{5/2}~~,
\end{eqnarray}
so that for the planet with initial velocity $v_{\rm p0}=6.2\times10^7$~cm~s$^{-1}$
(which is the escape velocity of the sun),
the Mach number is
\begin{equation}
{\cal M}=\frac{v_p(z)}{c_{s*}(z)}
\approx 5 \left(\frac{z}{10^{10}{\rm cm}}\right)^{-5/2}
\exp \left[-\left(\frac{z}{z_s}\right)^3\right]~~,
\end{equation}
Therefore, the planet motion inside the star is mildly supersonic and
a bow shock is formed. 
As a result, the planet mass decreases due to ablation.
The ram pressure of the stellar gas also causes lateral expansion of the planet,
which increases the cross sectional area $\sigma_{\rm p}$ 
[see Eq.~(\ref{eq:dragforce})].
It is difficult to treat these effects analytically, but they effectively increase
the value of $\eta$, so that $z_s$ becomes smaller.
Planets could also inflate as they approach the star due to tidal heating.
Recent observations have shown that hot Jupiters have a larger radius than
expected for given planet mass \citep[e.g][]{Baraffe2014}.
Taking into account all of these additional effects beyond our simple estimate 
in Eq.~(\ref{eq:stopping}), the actual stopping depth may be a factor of a few
smaller, say $z_s\approx 1-2\times10^{10}$~cm.
This is of order $r_{\rm J}\approx0.1 R_{\sun}$,
in agreement with the previous simple estimate given by
Eq.~(\ref{eq:eratio}) in \S~\ref{subsec:collision}.
Around this depth, the planet\rq{}s kinetic energy is suddenly released, so that
the thermal energy, $E_{\rm th,i}$, is deposited there.
The hot bubble arises  there. 
The bubble temperature is $T_i\sim10^{6-7}$~K, 
based on Eq.~(\ref{eq:initialT}), and  is slightly larger than
the temperature of the stellar gas at $z_s$, $T_*(z_s)$.

Following the collision, the expanding bubble rises
and finally escapes the stellar surface.
The expansion velocity is of order  the sound speed of the bubble,
$\sqrt{k_BT_i/\mu_p}$, which is of the order  the sound speed of the stellar gas
at the base, $c_{s*}(z_s)$.
The expansion time of the bubble  is $t_{\rm exp}(z)\sim z/c_{s*}(z_s)$.
This timescale is comparable to  the time that the stellar gas material fills
the rarefied region behind the planet,
 $t_{\rm cl}(z)\sim r_{\rm pl}/c_{s*}(z)$.
Hence, we obtain
$t_{\rm exp}(z)/t_{\rm cl}(z)\sim(z/r_{\rm pl})[c_{s*}(z)/c_{s*}(z_s)]
=(z_s/r_{\rm pl})(z/z_s)^{7/2}$.
Since $z_s$ is slightly larger than $r_{\rm pl}$, the rarefied region closes before the bubble expands.
However, a strong pressure wave will travel to the surface of the star and 
lift material from there out of the gravitational potential well of the star. 
The actual impact of a collision can only be reliably calculated with a numerical 
hydrodynamics simulation, which we leave for future work.

%
\section{Prompt Emission from expanding plasma bubble}
%
As seen in \S~\ref{sec:collision}, the collision between a star and 
a Jovian planet
releases  thermal energy $E_{\rm th,i}$ into a volume of 
radius $R_{\rm i}$ over a short time.
We approximate $R_{\rm i}\sim r_{\rm pl}$ and the number density of the confined gas,
$n_{\rm i} \sim3m_{\rm pl}/4\pi \mu_{\rm p}R_{\rm i}^3$, where $\mu_{\rm p}$ is the proton mass.
The plasma bubble is optically thick to its thermal photons since
initial optical depth is estimated as 
$\tau_{\rm i}=n_{\rm i}\sigma_{\rm T} R_{\rm i} = 3.7\times10^9(m_{\rm pl}/m_{\rm J})(R_{\rm i}/r_{\rm J})^{-2}$,
where $\sigma_{\rm T}$ is the Thomson cross section. 
Subsequently, the bubble expands due to its thermal pressure. 
Here, we focus on a simple estimate of the 
luminosity of the resulting emission by the bubble.
For simplicity, we assume that the 
gas has  uniform density and temperature, and a homologous
velocity profile.

The bubble is matter dominated, so that the initial temperature is given by
\begin{equation}
T_{\rm i}\sim\frac{GM_*\mu_{\rm p}}{3k_{\rm B}R_*}
=7.7\times10^6\xi_*~{\rm K}~~.
\label{eq:initialT}
\end{equation}
The initial radiation energy, 
$E_{\rm rad,i}\sim(aT_{\rm i}^4)(4\pi R_{\rm i}^3/3)\sim3.8\times10^{43}\xi_*^4(R_{\rm i}/r_{\rm J})^3$erg,
where $a$ is the radiation energy constant, is much smaller than $E_{\rm th,i}$
in Eq.~(\ref{eq:initialEth}).
The material is initially opaque to
its own thermal photons, allowing  very little internal energy to escape
 from its surface. The hot plasma expands adiabatically
with an expansion speed comparable to the escape velocity of the star,
$v_{\rm esc,*}= (2GM_*/R_*)^{1/2}=6.2\times10^7\xi_*^{1/2}$cm~s$^{-1}$.
The temperature declines adiabatically  as a function of  radius $R$, as 
$T(R)=T_{\rm i} (R/R_{i})^{-2}$, 
while the gas number density is given by 
$n(R)=n_{\rm i}(R/R_{\rm i})^{-3}$.

The expanding bubble becomes optically thin 
when it cools below the hydrogen recombination temperature of $\sim10^4$K.
The photosphere radius $R_{\rm ph}$ is determined by the condition that 
the photon diffusion time $t_{\rm diff}=n_{\rm e}\sigma_{\rm T}R^2/c$,
where $n_{\rm e}=n_{\rm e}(R)$ is the number density of free electrons,
is equal to the expansion time $t_{\rm exp}=R/v_{\rm esc,*}$ at $R_{\rm ph}$.
We define $x=n_{\rm e}/n(R)$ as the ionization degree at radius $R$.
The Saha equation for hydrogen,
\begin{equation}
\frac{1-x}{x^2}=n(R)\left(\frac{h^2}{2\pi \mu_{\rm e} kT(R)}\right)^{3/2}
\exp\left(\frac{13.6~{\rm eV}}{kT(R)}\right)~~,
\label{eq:saha}
\end{equation}
combined with $n(R)/n_{\rm i}=[T(R)/T_{\rm i}]^{3/2}$ and 
$t_{\rm diff}/t_{\rm exp}=x\tau_{\rm i}(T/T_{\rm i})(v_{\rm esc,*}/c)=1$,
can provide the equation for the temperature $T_{\rm ph}$ at the photosphere,
to numerically find $T_{\rm ph}\approx7200$~K.
Due to the exponential term in Eq.~(\ref{eq:saha}), the value of $T_{\rm ph}$ is
almost independent of the initial state of the bubble.
The photosphere radius is therefore,
\begin{eqnarray}
R_{\rm ph}&=&R_{\rm i} \left(\frac{T_{\rm i}}{T_{\rm ph}}\right)^{1/2}\nonumber\\
&=&2.3\times10^{11}\xi_*^{1/2}
\left(\frac{R_{\rm i}}{r_{\rm J}}\right)
\left(\frac{T_{\rm ph}}{7200~{\rm K}}\right)^{-1/2}~{\rm cm}~~,
\label{eq:photosphere}
\end{eqnarray}
and the ionization degree at $R=R_{\rm ph}$ is
$x(R_{\rm ph})=1.4\times10^{-4}(m_{\rm pl}/m_{\rm J})^{-1}(R_{\rm i}/r_{\rm J})^2\xi_*^{-1/2}(T_{\rm ph}/7200~{\rm K})^{-1}$.
The observer would detect  blackbody radiation with a temperature $T_{\rm ph}$
and a peak bolometric luminosity,
\begin{eqnarray}
L_{\rm p} &=&4\pi R_{\rm ph}^2\sigma T_{\rm ph}^4\nonumber\\
&=& 1.0\times10^{35} \xi_*
\left(\frac{R_{\rm i}}{r_{\rm J}}\right)^2
\left(\frac{T_{\rm ph}}{7200~{\rm K}}\right)^{3}~{\rm erg~s}^{-1}~~,
\label{eq:Lp}
\end{eqnarray}
where $\sigma$ is the Stefan-Boltzmann constant.
The peak flux density at frequency $\nu=\nu_{14}\times10^{14}$Hz and a distance 
$d=1d_{{\rm kpc}}$~kpc from the source is then
\begin{eqnarray}
F_\nu^{\rm p} &=&\frac{L_{\rm p}}{4\pi d^2}
\frac{15}{\pi^4\nu}\left(\frac{h\nu}{kT_{\rm ph}}\right)^3
f\left(\frac{h\nu}{kT_{\rm ph}}\right)\nonumber\\
&=& 38~\xi_*
\left(\frac{R_{\rm i}}{r_{\rm J}}\right)^2
\nu_{14}^2d_{{\rm kpc}}^{-2}
f\left(\frac{h\nu}{kT_{\rm ph}}\right)~{\rm mJy}~~,
\label{eq:flux_prompt}
\end{eqnarray}
where $f(y)=y(e^y-1)^{-1}$.
Table~1 provides  the flux density in various observation bands.
Note that since $\xi_*\approx1$ for $M_*\la 2M_{\sun}$, the observed flux 
hardly depends on stellar properties. 
Note that in deriving Eqs.~(\ref{eq:Lp}), (\ref{eq:flux_prompt}) and values in Table~1,
we assume that the gas initially expands adiabatically, that is
$T(R)\propto [n(R)]^{2/3}$.  This relationship is only valid for a monatomic gas of a fixed
ionization state.  The process of recombination releases energy,
which keeps the gas more isothermal than predicted by this relationship.  
This increases the photosphere radius from that estimated in 
Eq.~(\ref{eq:photosphere}), increasing the luminosity of the prompt emission.

The typical duration  of the transient is comparable to the dynamical time-scale,
\begin{equation}
\Delta T\sim \frac{R_{\rm ph}}{v_{\rm esc,*}} 
=3.7\times10^3\left(\frac{R_{\rm i}}{r_{\rm J}}\right)
\left(\frac{T_{\rm ph}}{7200~{\rm K}}\right)^{-1/2}{\rm s}~~,
\end{equation}
so that the total emission energy is
\begin{equation}
E_{\rm rad}\sim L_{\rm p} \Delta T =3.7\times10^{38}
\xi_*
\left(\frac{R_{\rm i}}{r_{\rm J}}\right)^3
\left(\frac{T_{\rm ph}}{7200~{\rm K}}\right)^{5/2}{\rm erg}~~,
\end{equation}
which is much smaller than the initial internal energy of the bubble.
Hence,  almost all the initial energy transforms to  kinetic energy
and gets dissipated when the expanding material interacts with the CSM.

\begin{table}
\label{table:optical}
\caption{Predicted optical/infrared peak flux density of the prompt emission.
The unabsorbed observed peak flux, $F_\nu^{\rm p}$, is for the distance $d=10$~kpc from the source with
$\xi_*=1$, $R_{\rm i}=r_{\rm J}$, and $T_{\rm ph}=7200$~K.}
\begin{tabular}{@{}lccc@{}}
\hline
Filter & $\lambda$
& $\nu$ & $F_\nu^{\rm p}$ \\
& [nm] &  [$10^{14}{\rm Hz}$] &  [mJy] \\
\hline
\hline
g\rq{} & 475 & 6.3 & 0.97  \\
r\rq{} & 622 & 4.8 & 1.2  \\
i & 763       & 3.9 & 1.2  \\
y & 1020   & 2.9 & 1.1  \\
J &1220    & 2.5 & 0.91  \\
H & 1630 & 1.8 & 0.66  \\
K & 2190 & 1.4 & 0.44  \\
L & 3450 & 0.87 & 0.21  \\
M & 4750 & 0.63 & 0.12  \\
N & 10500 & 0.29 & 0.028  \\
\hline
\end{tabular}
\end{table}

At the moment of tidal disruption, the planet is expected to be
vertically collapsed \citep[e.g.,][]{Kobayashi2004,Guillochon2009}.
The work done by the tidal force from the star is estimated to be
$W\sim(GM_*m_{\rm pl}/R_{\rm p}^2)(r_{\rm pl}/R_{\rm p})(r_{\rm pl}/2)
\sim\beta^3Gm_{\rm pl}^2/2r_{\rm pl}$.
If the collapsed matter is thermalized and half of this energy  is released, 
then the initial thermal energy becomes
$E_{\rm th,i}\sim8.6\times10^{42}\beta^3(m_{\rm pl}/m_{\rm J})^{5/3}$~erg,
where we use an approximate relation,
$(r_{\rm pl}/r_{\rm J})\approx(m_{\rm pl}/m_{\rm J})^{1/3}$  \citep{Rappaport2013}.
This is about two orders of magnitude smaller than the energy considered 
in Eq.~(\ref{eq:initialEth}).
The initial temperature is then estimated as
$T_{\rm i}\sim3.7\times10^4\beta^3(m_{\rm pl}/m_{\rm J})^{2/3}$K, which is of order
the recombination temperature.
The bubble expands at a speed comparable to the
free-fall velocity of the planet, 
$v_{\rm ff,pl}\sim(Gm_{\rm pl}/r_{\rm pl})^{1/2}=4.3\times10^6(m_{\rm pl}/m_{\rm J})^{1/2}(r_{\rm pl}/r_{\rm J})^{-1/2}$cm~s$^{-1}$.
Similarly to the previous calculation, we derive a temperature 
of $T_{\rm ph}\approx8100$~K at the photosphere radius 
$R_{\rm ph}=1.5\times10^{10}(m_{\rm pl}/m_{\rm J})^{2/3}\beta^{3/2}(T_{\rm ph}/8100~{\rm K})^{-1/2}$.
The peak bolometric luminosity is then 
$L_{\rm p}\sim6.7\times10^{32}\beta^{3}(m_{\rm pl}/m_{\rm J})^{4/3}(T_{\rm ph}/8100~{\rm K})^{3}$erg~s$^{-1}$.
This is smaller than the main prompt emission (see Eq.~5), however,
it could be larger for larger $\beta$ and/or $m_{\rm pl}$, 
in which case the emission could be detectable as a precursor arising
before the prompt emission.


\section{Radio afterglow}

The expanding plasma maintains a constant velocity $v_{\rm esc,*}$ out to the
deceleration radius,
\begin{equation}
R_{\rm dec}=\left(\frac{3m_{\rm pl}}{4\pi \mu_{\rm p}n_{\rm c}}\right)^{1/3}
=6.5\times10^{17}n_{\rm c}^{-1/3}
\left(\frac{m_{\rm pl}}{m_{\rm J}}\right)^{1/3}
{\rm cm}~~,
\end{equation}
where $n_{\rm c}$ is the density of the CSM.
The bubble reaches this radius after
$t_{\rm dec}=R_{\rm dec}/v_{\rm esc,*}=3.3\times10^2\xi_*^{-1}n_{\rm c}^{-1/3}
(m_{\rm pl}/m_{\rm J})^{1/3}$yr.
During the expansion, the flow interacts with the CSM, generating 
an external shock with
a Mach number of $\sim60$.
Electron acceleration at the shock results in radio synchrotron emission.
The emission lasts until the shock velocity declines to  
$\sim1\times10^7$~cm~s$^{-1}$,
below which the ionization  of the acceleration region drops rapidly
\citep{Shull1979}, 
so that wave damping due to collisions with neutral atoms 
prevents electrons from being accelerated at the shock front
\citep{Drury1996,Bykov2000}.
Assuming the Sedov solution after a time $t_{\rm dec}$ 
($R\propto t^{2/5}$ and $v\propto t^{-3/5}$), 
we estimate the epoch $t_{\rm end}$ at which particle acceleration ceases to be
$t_{\rm end}\sim6.9\times10^3\xi_*^{-1/6}n_{\rm c}^{-1/3}(m_{\rm pl}/m_{\rm J})^{1/3}$yr,
corresponding to the radius
$R_{\rm end}\sim2.2\times10^{18}\xi_*^{1/3}n_{\rm c}^{-1/3}(m_{\rm pl}/m_{\rm J})^{1/3}$cm.
When the acceleration stops, the high-energy electrons starts to escape from
the shocked region with an escape time of $\sim1$--10~yr, resulting in
rapid fading of the emission.

Next we provide  a simple estimate of the observed flux and surface brightness of the
radio synchrotron emission from the expanding bubble of radius $R$.
The total number of nonthermal electrons, $N_{\rm e}$, is a fraction $\eta_{\rm e}$ of
the number of particles originating  from the upstream region of the shock over
a dynamical time $t_{\rm dyn}=R/v$, where $v\sim v_{\rm esc,*}$ is the expansion speed,
that is, 
$N_{\rm e} =\eta_{\rm e}(4\pi R^2n_{\rm c}v)t_{\rm dyn}=4\pi \eta_{\rm e} n_{\rm c}R^3$.
We assume a single power-law form of the nonthermal  electron distribution,
$N(\gamma) \propto \gamma^{-p}$, for $\gamma_{\rm m}<\gamma<\gamma_{\rm M}$,
where $\gamma$ is the electron Lorentz factor.
If the dynamical time is sufficiently long ($t_{\rm dyn}\ga2\times10^3$yr), 
the maximum Lorentz factor $\gamma_{\rm M}$ is determined by the balance of
the acceleration time and the synchrotron cooling time 
\citep[e.g.,][]{Yamazaki2004,Yamazaki2006,Yamazaki2015},
yielding
$\gamma_{\rm M}\approx1\times10^8 (\xi_*/fB_{-5})^{1/2}(v/v_{\rm esc,*})$, 
where $B_{-5}=(B/10~\mu{\rm G})$ is the post-shock magnetic field strength, 
and $f$ is a numerical factor of order unity which is
determined by the properties of scattering waves and shock geometry at the acceleration site.
Thus, we find that radio-emitting electrons have a much smaller Lorentz factor than $\gamma_{\rm M}$.

The  synchrotron cooling time of radio emitting electrons 
 is much longer than the dynamical time. Hence, the observed flux density at 
frequency $\nu$ is 
\citep[e.g.,][]{Sari1998}
\begin{eqnarray}
F_\nu&\sim&\frac{N_{\rm e}}{4\pi d^2}\frac{\mu_{\rm e}c^2\sigma_{\rm T}B}{3e}
\left(\frac{\nu}{\nu_{\rm m}}\right)^{(1-p)/2}\nonumber\\
&=& 4.0\times10^2
\frac{n_{\rm c} \eta_{-5}B_{-5}}{d_{\rm kpc}^2}
\left(\frac{R}{10^{18}{\rm cm}}\right)^3
\left(\frac{\nu}{\nu_{\rm m}}\right)^{(1-p)/2}
{\rm Jy}~~,\nonumber\\
&&
\label{eq:Fnu}
\end{eqnarray}
where $\eta_{-5}=(\eta_{\rm e}/10^{-5})$ and
$\nu_{\rm m}=28B_{-5}\gamma_{\rm m}^2$Hz is the characteristic synchrotron 
frequency from electrons with a minimum Lorentz factor $\gamma_{\rm m}$.
Note that for our parameters, $(\nu/\nu_{\rm m})^{(1-p)/2}$ is much lower than unity
(see Table~2).
The surface brightness, $S_\nu$, is the flux density divided by the solid angle of the source,
$\Omega\sim\pi(R/d)^2$, yielding
\begin{equation}
S_\nu\sim1.2\times10^9n_{\rm c}
\eta_{-5}B_{-5}
\frac{R}{10^{18}{\rm cm}}
\left(\frac{\nu}{\nu_{\rm m}}\right)^{(1-p)/2}
{\rm Jy~sr}^{-1}~~.
\label{eq:Snu}
\end{equation}
Table~2 shows the results for different values of $p$ with fixed parameters,
$n_{\rm c}=\eta_{-5}=B_{-5}=\gamma_{\rm m}=1$.
For  comparison, the surface brightness of the faintest Galactic supernova remnants
is $\sim10^{4}$Jy~sr$^{-1}$ \citep{Arbutina2005}, 
of the same order as the typical value of 
the diffuse Galactic radio emission \citep[e.g.,][]{deOliveira2008}.

\begin{table}
\label{table:radio}
\caption{Radio surface brightness at a radius $R=10^{18}$cm and frequency 
$\nu=1$~GHz for power-law electron distribution with index $p$.
Other parameters are taken as
$n_{\rm c}=\eta_{-5}=B_{-5}=\gamma_{\rm m}=1$.}
\begin{tabular}{@{}lcc@{}}
\hline
$p$ & $(\nu/\nu_{\rm m})^{(1-p)/2}$
& $S_{\nu=1{\rm GHz}}$ \\
&  &  [Jy~sr$^{-1}$] \\
\hline
\hline
2.0  &    $1.7\times10^{-4}$    &   $2.0\times10^5$  \\
2.1  &    $7.0\times10^{-5}$    &   $8.4\times10^4$  \\
2.2  &    $2.9\times10^{-5}$    &   $3.5\times10^4$  \\
2.3  &    $1.2\times10^{-5}$    &   $1.5\times10^4$  \\
2.4  &    $5.2\times10^{-6}$    &   $6.2\times10^3$  \\
2.5  &    $2.2\times10^{-6}$    &   $2.6\times10^3$  \\
3.0  &    $2.8\times10^{-8}$    &   $34$  \\
%
\hline
\end{tabular}
\end{table}


\section{Discussion}
\label{sec:summary}

\subsection{Event rate}

The event rate of the transients considered in this paper is highly uncertain.
It depends on various processes such as
the formation and early dynamics of giant planets.
The rate per galaxy is  roughly given by
$\sim{\rm SFR}\times (\alpha N_{\rm pl})/\langle m\rangle 
\approx5(\alpha N_{\rm pl})$~yr$^{-1}$,
where SFR is the star formation rate (see \S~1), 
$N_{\rm pl}$ is the average number of giant planets per star,
and $\alpha$ is the fraction of planets that are close enough to directly hit the stellar surface.
At present, it is highly uncertain how many giant planets are formed in 
a circumstellar disk because of the unknown disk mass.
Recent numerical simulations have indicated that disks are likely to be very massive
so as to harbor a lot of giant planets, suggesting
$N_{\rm pl}$ could be $\sim10$ \citep[e.g.,][]{Machida2011}.
The typical value of $\alpha$ is also uncertain.
To directly impact the stellar surface, planets must have a penetration factor larger than unity
(see \S~\ref{sec:ingestion}). 
It is possible that planet-planet interactions increase the eccentricity
making the pericenter small enough for a direct hit on the star \citep{Li2014}.
However,  tidal dissipation would tend to limit the eccentricity growth 
when the pericenter becomes small.
Another regime allowing for a direct collision between the planet and the star 
emerges in the early phase of the circumstellar disk, when the cloud core is still collapsing
and the motion of the newly formed planet is chaotic.
Some planets are formed at a large radii ($\sim20$--50~AU) and orbit only for
2--3 orbital periods, after which they fall into the star \citep{Machida2011}.
In this case, the eccentricity of the final orbit before the impact on the star may be near unity.
Here we expect that $\alpha N_{\rm pl}$ would be of order unity or even larger.
Future observation of the rate of the transients will constrain these highly
uncertain scenarios and shed light on  the process of the planet formation.
In particular, if the event rate is high,
we will be able to confirm that the circumstellar disk is massive.

It is also possible that after ingestion of Jovian planets, stars become metal rich
\citep[e.g.,][]{Sandquist1998,Sandquist2002,Cody2005}. 
One can get an upper limit on the event rate from the fraction of metal rich stars. 
However, the high metallicities could result from other processes 
(e.g. inhomogeneous supernova enrichment of the interstellar medium). 
The actual rate can be calibrated as a fraction of this maximum rate.

\subsection{Prospects for future observations}

While the prompt emission flare from the tidal disruption of a planet
cannot be detected in the optical  band due to  Galactic dust extinction,
the unabsorbed flux in the K-band  is $\sim0.2$~mJy at the distance of 10~kpc,
potentially detectable with infrared sky surveys, such as
UKIRT Infrared Deep Sky Survey \citep[UKIDSS:][]{Hewett2006,Lawrence2007} and
VISTA Variables in the Via Lactea \citep[VVV:][]{Minniti2010}.
If $\alpha N_{\rm pl}\la1$, the expected event
 rate  $\sim5(\alpha N_{\rm pl})$~events~yr$^{-1}$ 
for the entire Galactic plane (see \S~1),  is too small for detectability
by current transient surveys.
The duration $\Delta T$ of the expected prompt flares could be comparable to that of superflares
of stars \citep[e.g.,][]{Schaefer2000,Maehara2012}. 
The total emission energy in the
optical band is about an order of magnitude larger than the largest
superflares \citep{Shibayama2013,Balona2015}.

The Andromeda galaxy,
M31, located at the distance of 0.78~Mpc \citep{Stanek1998}, has a similar mass
 and star formation rate
to  the Milky Way \citep{Williams2003}, and hence a  similar event rate of the transients
proposed in this paper.
Unlike the Milky Way case, telescopes with a large field of view can cover 
the entire volume of M31.
The expected AB~magnitudes are 25.9 and 25.7~mag at g and r~bands, respectively,
for our fiducial parameters. The fluxes become higher for planets more larger than Jupiter.
Using the pixel lensing technique  
\citep[i.e.,  differencial image photometry:][]{Crotts1992,Baillon1993,Tomaney1996,CalchiNovati2010},
those transients could be detected by future instruments with better sensitivity like Subaru
Hyper Suprime-Cam\footnote{
http://www.naoj.org/Projects/HSC/index.html
}.

Events in the Large Magellanic Cloud (LMC) could also be detected.
For the distance of 48.5~kpc to LMC \citep{Macri2006}, the
expected AB~magnitudes are 19.9 and 19.6~mag at g and r~bands respectively
for our fiducial parameters, which are detectable by current surveys like
PAndromeda \citep{Lee2012}.
The event rate is only slightly smaller than M31 because the total 
star formation rate of the LMC is $\approx0.4 M_{\sun}$~yr$^{-1}$ \citep{Harris2009}.

If the electron index $p$ is smaller than about 2.3, the radio afterglow would be detectable
for $\sim10^{3-4}$yr after the disruption event.
The radio surface brightness is expected to be lower than young supernova remnants
 because the magnetic field is weak.
As a result, the surface brightness and diameter of the source  
would be distinguishable from
the values  expected for supernova remnants \citep{Arbutina2005}. 
Note that the classical nova eruption ejects typically $10^{-5}$ to $10^{-4}M_{\sun}$
of matter \citep[e.g.,][]{Bode2010, Roy2012}, 
which is one or two orders of magnitude smaller than that assumed in our
present model, $m_{\rm pl}\sim10^{-3}M_{\sun}$.
The observed radio emission of novae is typically well characterized by the free-free
emission process, which peaks at $\sim1$~yr after the eruption \citep[e.g.,][]{Ribeiro2014}.
After that, the observed radio emission declines, and no radio synchrotron halo has been
ever detected \citep[e.g.,][]{Ribeiro2014}.
Since the ejecta mass of novae and our planet-star collision model are close,
the current radio upper limits on the synchrotron shock emission may potentially constrain
the present model.
Equations~(\ref{eq:Fnu}) and (\ref{eq:Snu}) imply that the flux $F_\nu$ and the
surface brightness $S_\nu$
increase until the deceleration time $t_{\rm dec}$,
and takes maximum at $R=R_{\rm dec}$ if other model parameters are constant with time.
Since $R_{\rm dec}\propto M_{\rm ej}^{1/3}$ where $M_{\rm ej}$ is the ejecta mass
(that is, $M_{\rm ej}\approx m_{\rm pl}\sim10^{-3}M_{\sun}$ for our present model, and
$M_{\rm ej}\sim10^{-5}-10^{-4}M_{\sun}$ for the nova eruption), we get
$F_{\nu}\propto R_{\rm dec}^3\propto M_{\rm ej}$ and
$S_{\nu}\propto R_{\rm dec}\propto M_{\rm ej}^{1/3}$,
so that the radio synchrotron afterglows of the classical novae show surface brightness
several times smaller than those of the star-planet collision.
As shown in Table~2, our present synchrotron shock model predicts low surface brightness 
that is comparable to the present-day detection limit.
This may explain why the related radio synchrotron afterglow has not 
been detected as of yet.


\section*{Acknowledgments}

We thank 
Edo~Berger,
Shu-ichiro~Inutsuka,
Yutaka~Ohira,
Lorenzo~Sironi,
Masaomi~Tanaka,
Paola~Testa
and
Makoto~Uemura
for helpful discussion and valuable comments. 
We also thank the anonymous referee for valuable comments to improve the paper.
This work was supported in part by grant-in-aid 
from the Ministry of Education, Culture, Sports, 
Science, and Technology (MEXT) of Japan, 
No.~15K05088 (R.~Y.),
research grant of the Chungbuk National University in 2015 (K.~H.), 
and NSF grant AST-1312034 (A.~L.).
R.~Y. also thank ISSI (Bern) for support of the team 
``Physics of the Injection of Particle Acceleration at 
Astrophysical, Heliospheric, and Laboratory Collisionless Shocks''. 





\label{lastpage}
\end{document}